\documentclass[fleqn,twoside]{article}
\usepackage{espcrc2}

\usepackage{graphicx}
\usepackage[figuresright]{rotating}

\newcommand{\AmS}{{\protect\the\textfont2
  A\kern-.1667em\lower.5ex\hbox{M}\kern-.125emS}}

\title{The local structure of topological charge fluctuations in QCD}

\author{I.~Horv\'ath\address[UK]{Department of Physics \& Astronomy, 
        University of Kentucky, Lexington, KY 40506, USA},
        S.J.~Dong\addressmark[UK],
        T.~Draper\addressmark[UK],
        F.X.~Lee\address{Center for Nuclear Studies, 
        George Washington University, Washington, DC 20052, USA}\address[JL]
        {Jefferson Lab, 12000 Jefferson Avenue, Newport News, VA 23606, USA},
        K.F.~Liu\addressmark[UK],
        J.B.~Zhang\address{CSSM and Dept. of Physics and Math. Physics, 
        University of Adelaide, Adelaide, SA 5005, Australia},
        H.B.~Thacker\address[UVA]{Department of Physics, University of Virginia,
        Charlottesville, VA 22901, USA}}
       
\begin{document}

\begin{abstract}
We introduce the Dirac eigenmode filtering of topological charge density associated 
with Ginsparg-Wilson fermions as a tool to investigate the local structure of topological 
charge fluctuations in QCD. The resulting framework is used to demonstrate that the bulk 
of topological charge in QCD does not appear in the form of unit quantized lumps. This 
means that the mixing of ``would-be'' zeromodes associated with such lumps is probably 
not the prevalent microscopic mechanism for spontaneous chiral symmetry breaking in QCD. 
To characterize the coherent local behavior in topological charge density at low energy, 
we compute the charges contained in maximal coherent spheres enclosing non-overlapping 
peaks. We find a continuous distribution essentially ending at $\approx 0.5$. Finally, 
we study, for the first time, the overlap-operator topological-charge-density correlators 
and find consistency with non-positivity at nonzero physical distance. This represents 
a non-trivial check on the locality ({\em in gauge paths}) of the overlap Dirac operator 
for realistic gauge backgrounds.    
\vspace{1pc}
\end{abstract}

\maketitle
\input epsf

The aim of this work is to set up a physically meaningful and unbiased framework
to probe the low-energy structure of topological charge fluctuations in QCD,
and use it to study some related physics~\cite{Hor02B}. As a starting point, 
we will consider the following question. Can one confirm or disprove 
a qualitative picture of vacuum fluctuations where {\em most} topological 
charge appears in the form of identifiable coherent unit quantized lumps? This 
question is interesting because if the answer is yes, then one could possibly 
argue that S$\chi$SB, the $\theta$-dependence, and the $\eta'$ mass have common 
qualitative origin modeled in the instanton picture. However, if the answer is no, 
then this frequently-used intuitive reasoning has no verifiable basis. 
The main reason why a {\em lumpy vacuum} (vacuum dominated by unit quantized lumps) 
represents a dividing line is that such vacuum has a very specific effect 
on fermions. In particular, the chiral ``would-be'' zeromodes $\chi^i$ implied 
by the index theorem and localized on the lumps ${\cal L}_i$ would mix and form 
a special subspace of states 
$\psi^i \,\approx\, \sum_j a^{ij}\,\chi^{j}\,,\; i,j\,=\, 1,\ldots N_L$
which will be referred to as the {\em topological subspace}.

The usual reasoning associated with topological mixing contains a subtlety
rooted in the fact that the Euclidean $q(x)$ is odd with respect to reflections
and consequently
$<\!\! q(x) q(0)\!\!> \,\,\le\, 0$ for $|x|>0$, \cite{SeSt} (see also~\cite{Tha02A}). 
This means that any four-dimensional coherent structure in the long-wavelength 
components of $q(x)$ will be obscured by anticorrelated background fluctuations. 
This fact appears to be at the heart of all the problems associated with studying 
topology locally and, since we discuss this in the continuum, it has nothing 
to do with lattice as is sometimes incorrectly argued. 

Since the regions ${\cal L}_i$ cannot be identified in $q(x)$, the topological 
mixing scenario is ill-defined (and thus can not be verified) until one {\em defines} 
the coherent lumps for realistic backgrounds in the first place. The issue here 
is that, naively, the topological mixing picture relates the local behavior of 
infrared modes, where the ultraviolet fluctuations of the underlying gauge field are 
filtered out~\cite{Hor01A,Hor02A}, to the behavior of $q(x)$ containing 
all ultraviolet fluctuations. This suggests that a logically consistent framework 
can be built around the possiblity of defining an {\em effective topological charge 
density}, based on the infrared eigenmode expansion. This can be conveniently
achieved starting from lattice topological charge density associated with GW 
fermions~\cite{Has98A}
\begin{equation}
          q_x \,=\, \frac{1}{2}  \mbox{\rm tr} \,\gamma_5 \,D_{x,x} \,=\, 
          -\mbox{\rm tr} \,\gamma_5 \, (1 - \frac{1}{2}D_{x,x})
\end{equation}
and defining the effective densities through~\cite{Hor02B}
\begin{equation}
    q_x^{(k)} \,\equiv\, 
    -\sum_{i=1}^{N_0} c^{0,i}_x\,
    -\sum_{j=1}^k (2 - \mbox{\rm Re}\, \lambda_j)\, c^{\lambda_j}_x 
\end{equation}
where $D \psi^{\lambda} = \lambda \psi^{\lambda}$, and 
$c^\lambda_x = \psi^{\lambda \,+}_x \gamma_5 \psi^{\lambda}_x$ is the local
chirality. $N_0$ zeromodes and $k$ conjugate complex eigenpairs (ordered by 
increasing real part of the eigenvalue) contribute to $q^{(k)}$.

Like the full density $q_x$, the filtered densities $q^{(k)}_x$ satisfy the exact 
index theorem~\cite{Has98A}, and yield identical global fluctuations since 
$\sum_x q_x^{(k)}=Q$, $\forall k$. However, the singular short-distance 
behavior of $q_x$ is smoothed out~\cite{Hor02B}, and the coherent structures 
in $q_x^{(k)}$ can potentially survive the continuum limit. The topological mixing 
picture can now be formulated {\it consistently} and checked. In particular, 
the picture asserts that for a typical gauge background, the lowest effective 
densities $q_x^{(k)}$ will be dominated by $N_L$ sign-coherent regions ${\cal L}_i$, 
and the topological charge contained in these regions will saturate at $k_L$ 
satisfying $N_L = N_0 + 2 k_L$, when
$\sum_{x\in {\cal L}_i} q_x^{(k_L)}\approx \pm 1$. Note that the existence of
unit lumps in effective densities would then be consistent with the basic 
assumption of topological mixing, according to which the low-lying modes 
are built of ``would-be'' zeromodes $\chi^i$ localized on ${\cal L}_i$ 
so that $\sum_{x\in {\cal L}_i} |\chi_x^i|^2 \approx 1$. Indeed, 
since $\lambda_i \approx 0$ for modes in the topological subspace, it follows that
\begin{equation}
  \sum_{x\in {\cal L}_i} \sum_{j=1}^{N_L}\, \psi_x^{j+}\gamma_5 \psi_x^j
  \,\approx\, 
  \sum_{x\in {\cal L}_i} q_x^{(k_L)}\approx \pm 1   
\end{equation}

We now test this proposition on QCD backgrounds. If the low-energy structure 
in a configuration with topological charge $Q$ is dominated by $|Q|+k_L$ 
coherent unit lumps (antilumps) and $k_L$ antilumps (lumps), then we should observe 
that $\sum_x |q^{(k_L)}_x|\equiv Q^{k_L,abs}\approx N_L = |Q|+2k_L$. In fact, the 
quantity $|Q| + 2k - Q^{k,abs}$ would be close to zero for all $k\le k_L$ 
(each mode in the topological subspace contributes $\approx 1$ to $Q^{k,abs}$), 
and it would start to increase rapidly for $k>k_L$ (modes outside the topological
subspace contribute $\ll 1$). If the topological mixing scenario was relevant, 
monitoring the $k$-dependence of $|Q| + 2k - Q^{k,abs}$ could thus serve 
as a procedure for determining the dimension of the topological subspace for 
a given configuration. We have computed this $k$-dependence using overlap Dirac 
operator for a few dozen of configurations at various lattice spacings, but have 
never observed the expected break at definite $k_L$. Instead, the typical robust 
behavior is shown below for an ensemble of six $12^4$ Wilson gauge configurations 
at $\beta=5.91$.

\medskip
\epsfxsize .85\hsize
\centerline{\epsffile {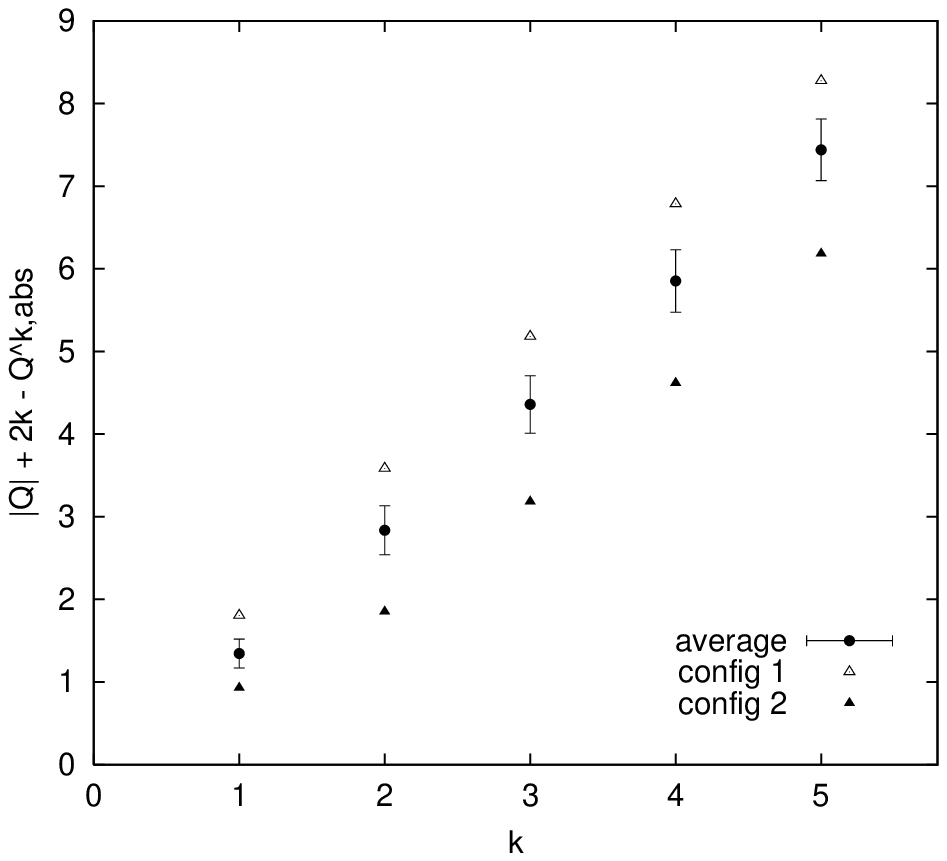}}
\medskip

\noindent Given our volume, $|Q| + 2k - Q^{k,abs}$ should be close to zero for 
$k \approx 1,2$, as the consistency with the value of topological susceptibility 
would require, but we observe large values instead. This apparent inconsistency leads 
us to conclude that the effective densities for QCD configurations {\it are not 
dominated by coherent unit lumps}.
  
An important feature of effective densities is that they can be used to study the 
low-energy structure of topological charge fluctuations {\it without prejudice} 
to a specific picture. As a first step towards this we calculate the typical amounts 
of topological charge contained in the maximal coherent spheres enclosing the peaks 
observed in effective densities. We fix the resolution for identification of such 
peaks by requiring that most of the resulting structures do not overlap~\cite{Hor02B}.
Shown below is the distribution of charges obtained from six $12^4$ configurations 
at $\beta=5.91$ using $q^{(9)}$.

\medskip
\epsfxsize .85\hsize
\centerline{\epsffile {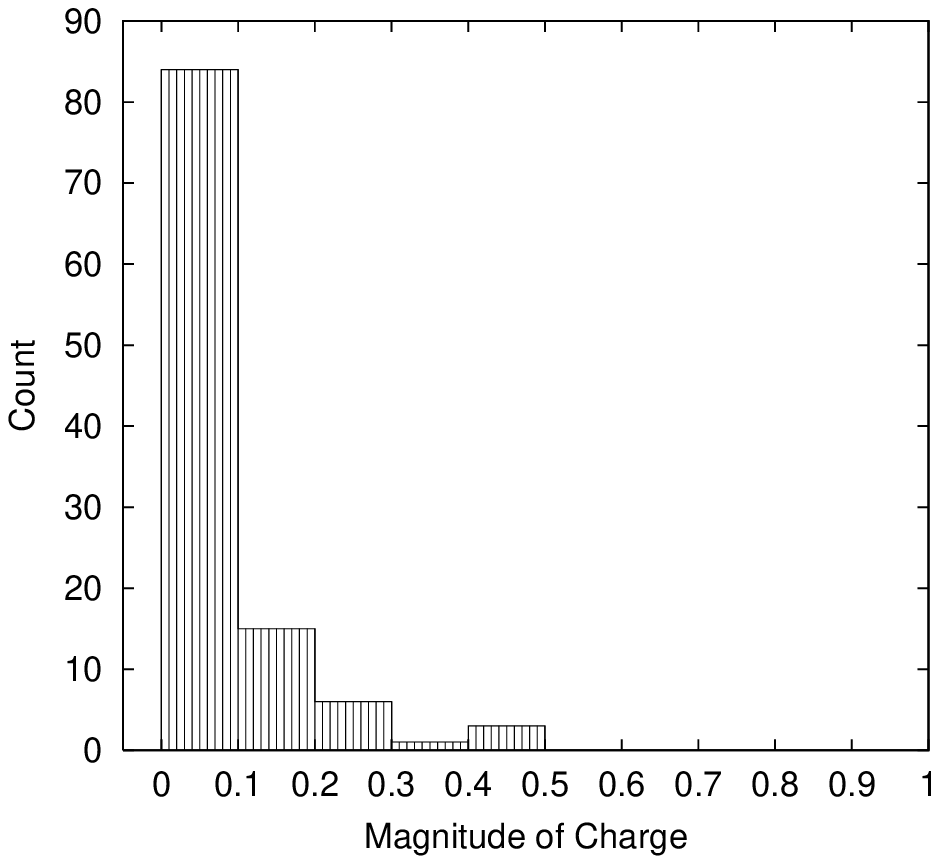}}
\medskip

\noindent The charges contained in the peaks are reasonably 
stable with respect to inclusion of additional eigenmodes. Interestingly, the distribution 
essentially ends at $0.5$ and most of the charge is carried by structures containing 
smaller amounts. This is consistent with the scenario emerging from the assumption 
of dominance by center vortices~\cite{EngRei00}.

\medskip
\epsfxsize .85\hsize
\centerline{\epsffile {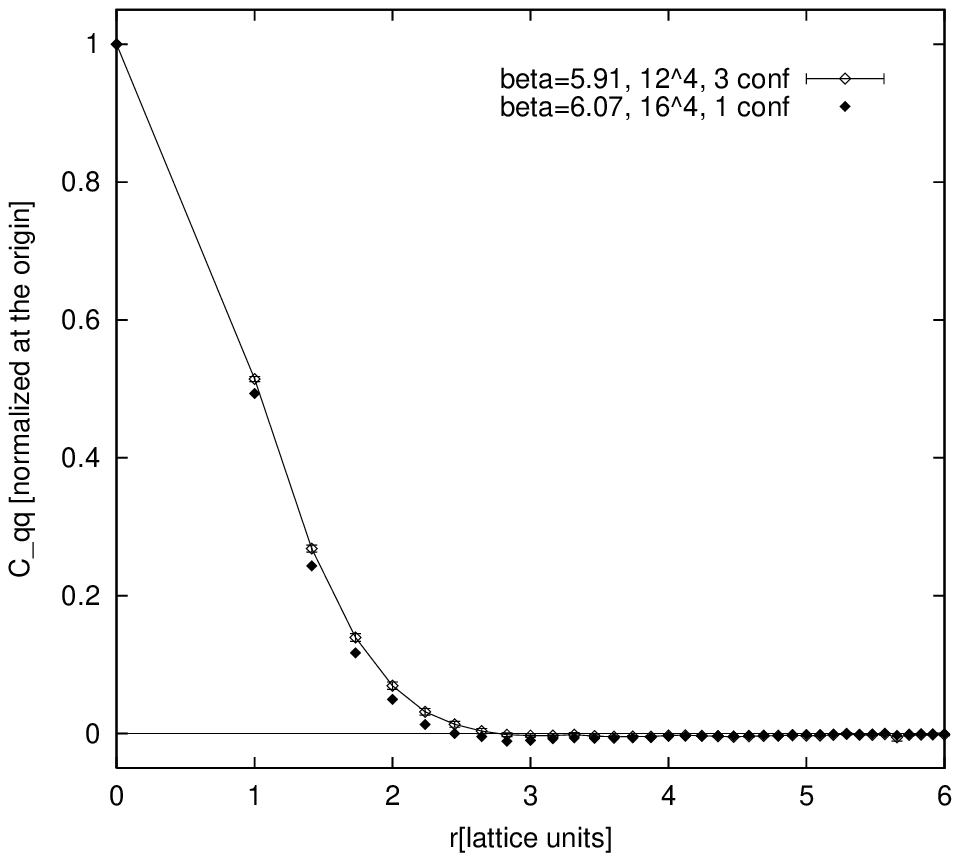}}
\medskip

Finally, we come back to the question of whether the overlap $q_x$ complies with 
non-positivity of the correlator at non-zero distances. For any valid 
reflection-positive lattice action $S(U)$ and well-defined {\em local} $q_x$, 
the consistency of the continuum limit requires that 
$\lim_{a\rightarrow 0, a|x|=r_p} < q_x q_0 > \;\le\, 0$, where $r_p$ is 
a non-zero physical distance. Since the Wilson gauge action is reflection positive, 
in our case this property relies on the locality of overlap $q_x$, and hence 
on the locality of $D_{x,x}$. This is not guaranteed since, contrary to the 
locality in fermionic variables, the locality in gauge paths has not been 
studied for realistic backgrounds. In the plot above we show preliminary results 
for $<q_x q_0>$ (normalized at the origin) at two lattice spacings. Even though 
the number of configurations is small, the behavior is very robust and reveals 
that the effect of {\it chiral smoothing}~\cite{Hor02A} is about two lattice 
spacings. This appears to be scale-independent. Indeed, the size of the positive 
core drops accordingly (see figure below), and is consistent with being a contact 
term in the continuum limit. This justifies the use of overlap $q_x$ and 
provides a nontrivial check on the locality properties of overlap Dirac operator. 
A detailed account of these observations is in preparation. 

\medskip
\epsfxsize .85\hsize
\centerline{\epsffile {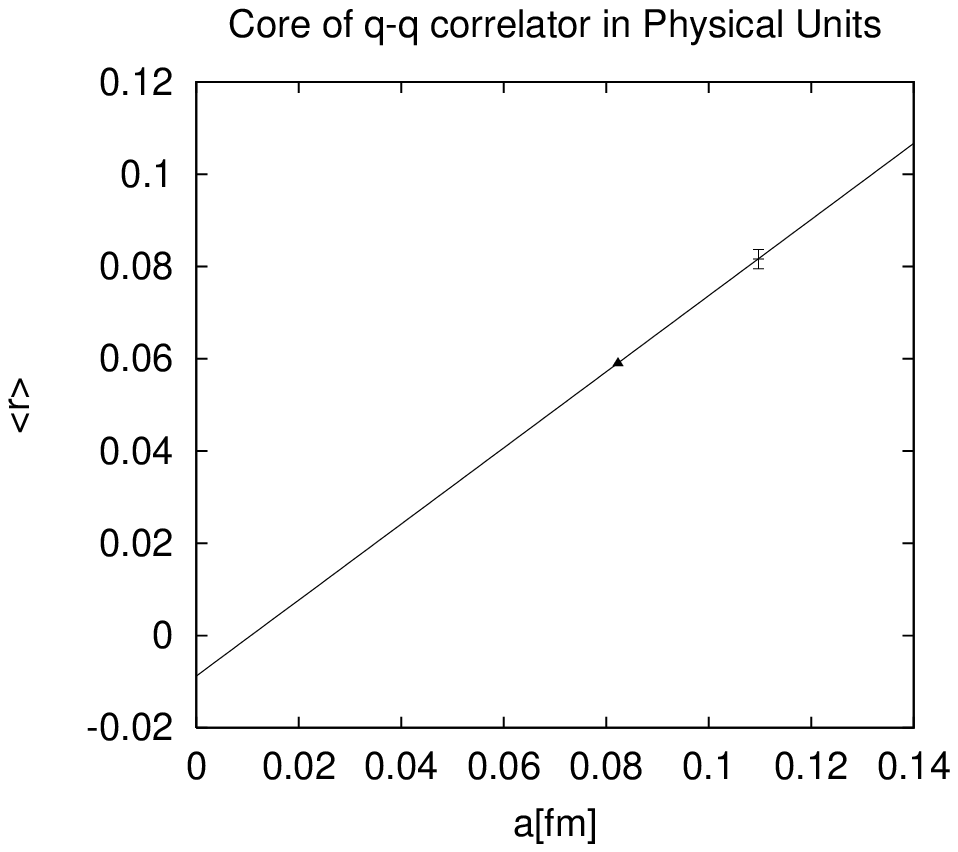}}

\end{document}